\documentstyle[multicol,prl,aps]{revtex}
\begin{document}

\title{
Exact finite-size spectrum for the multi-channel Kondo model
and Kac-Moody fusion rules
}

\author{Satoshi Fujimoto$^1$ and Norio Kawakami$^2$}

\address{
$^1$Department of Physics, Faculty of Science,
Kyoto University, Kyoto 606, Japan\\
$^2$Department of Material and Life Science,
and Department of Applied Physics, \\
Osaka University, Suita, Osaka 565, Japan
}
\maketitle
\begin{abstract}
The finite-size spectrum for the  multi-channel
Kondo model is derived analytically from the
exact solution, by mapping the nontrivial Z$_{n}$ part of the Kondo
scattering into that for the RSOS model coupled with the impurity.
The analysis is performed for the case of
$n-2S=1$, where $n$ is the number of channel and $S$
is the impurity spin. The result obtained is in accordance with
the Kac-Moody fusion hypothesis proposed by Affleck and Ludwig.
\end{abstract}
\begin{multicols}{2}
The multi-channel Kondo problem is one of the
most interesting issues in strongly correlated electron
systems\cite{multi,tw,ad,sch}.
It possesses the non-trivial fixed point characterized by a
non-Fermi liquid behavior for the
overscreening case. A possibility of its realization in some
Uranium compounds has been discussed extensively\cite{multi2,ks}.
Recently, Affleck and Ludwig developed a boundary conformal field
theory (CFT) approach to this problem, and derived various
non-trivial results for low-energy properties\cite{la}.
Their key proposal is the so-called fusion hypothesis;
i.e. when absorbing the impurity spin into conduction electrons,
the selection rule for spin quantum numbers is changed
following the Kac-Moody fusion rules.
Using this hypothesis they predicted the finite-size spectrum,
which plausibly yields the $\pi/2$ phase shift for $n\leq 2S$,
as well as  the non-Fermi liquid spectrum for $n>2S$,
where $n$ is the number of channel (flavor), and $S$
is the impurity spin.

On the other hand, the Bethe-ansatz solutions to various
Kondo models have been obtained\cite{exact,tw2},
which in principle can be compared with the
above prediction. This was performed
for the single channel Kondo model, confirming that the Bethe-ansatz
indeed gives the spectrum in accordance  with
the fusion hypothesis\cite{fks}.
For the overscreening model\cite{tw,ad}, however, the correct
finite-size spectrum has not been obtained analytically from the
Bethe-ansatz solution. The difficulty comes from that
the  multi-channel model relies on the string hypothesis
even for the ground state, which is valid only in the
thermodynamic limit\cite{tw,ad}. Hence the ordinary
Bethe-ansatz approach suffers from this pathology,
leaving the analytic computation
of the finite-size spectrum still open.

In this paper, we propose a way to deduce
the finite-size spectrum of the
overscreening Kondo model analytically from the
exact solution. A key idea is to
convert the nontrivial Z$_{n}$ part of the original
Kondo scattering into that for the critical restricted
solid-on-solid (RSOS) model coupled
with the impurity. The finite-size spectrum is then calculated
by the functional equation method based upon fusion hierarchy
\cite{res2,kp,kns}.
The obtained spectrum
plausibly reproduces the fusion hypothesis of Affleck and Ludwig.

We start by recapitulating the difficulty in the ordinary
Bethe-ansatz approach.  The multi-channel Kondo model
is given by
\begin{equation}
H = \sum_{k, l, \sigma} \epsilon_k c_{kl\sigma}^{\dagger}
 c_{kl \sigma}
    + J \sum_{k,k',l,\sigma,\sigma '} c_{kl\sigma}^{\dagger}
 (\sigma_
      {\sigma \sigma '} \cdot {\bf S}) c_{k'l\sigma '}
\end{equation}
with the antiferromagnetic coupling $J>0$,
where $l$ is the flavor index ($l=1, 2, \cdots, n$),
$\sigma= \uparrow, \downarrow$ is the spin index,
and other notations are standard. This model was solved by the
coordinate Bethe-ansatz method, which
has described the thermodynamics exactly\cite{tw,ad}.
However, a naive application of finite-size techniques
\cite{dw,woy} based on the string hypothesis turns out to fail
for the overscreening case, and gives only
the gaussian part of the spectrum  for the spin sector. Namely,
the total spectrum leads to a pathological form,
\begin{eqnarray}
E&=&\frac{2\pi v}{N}\biggl(\frac{(Q-n)^2}{4n}+\frac{S_z^2}{n}
+(\mbox{flavor part}) \nonumber \\
&&+n_Q+n_s+n_f\biggr),\label{eqn:fss1}
\end{eqnarray}
for the finite system with linear size $N$,
where  $v$ is the Fermi velocity, and
$n_Q$, $n_s$, and $n_f$ are non-negative integers.
Here the first term expresses the charge excitation
specified by the quantum number $Q$, and
the second is the spin excitation labeled
by the magnetization $S_z$.
Since the spin sector of the multi-channel Kondo model
may be correctly described by the
level-$n$ SU(2) Wess-Zumino-Witten model
with the central charge $c_{WZW}=3n/(2+n)$
\cite{la}, it is seen from (\ref{eqn:fss1})
that the $Z_n$ parafermion sector
with central charge $c=c_{WZW}-1=2(n-1)/(n+2)$
is lacking. This type of difficulty is common to
many integrable models with affine symmetry\cite{am}.

We thus resort to alternative techniques other than the
coordinate Bethe ansatz to recover the correct spectrum for
the lacking Z$_n$ parafermions.
Though it seems not easy to resolve this problem
completely, we propose an analytic way to investigate
the nontrivial $Z_n$ parafermion part.
For this purpose, we first exploit the following properties of
the S-matrix for ``physical particles'' \cite{res,zz,fend}
in the overscreening
Kondo model. The bulk S-matrix in the spin
sector is decomposed into two parts\cite{res,fend};
the S-matrix for the 6-vertex model
which is equivalent to the gaussian theory,
and that for the Z$_{n}$ model.
Since the Z$_{n}$ model can be described
in terms of the RSOS model\cite{abf,zf},
the corresponding S-matrix is given by the face weight
in the RSOS model \cite{res,ts}.
A remarkable point for the overscreening model is that
the interaction between ``physical particles''
and the impurity is described by the S-matrix
for multi-kinks, which is given by the fusion of the face
weights of the RSOS model with the
fusion level $p=n-2S$.
This has been previously noticed by Fendley
\cite{fend} and  Martins\cite{mar}.
Therefore, it is seen that {\it  nontrivial properties in the
overscreening model is essentially determined by
the RSOS model coupled with the impurity}.
Based on these observations, we now consider
the transfer matrix  $T(u)$ for the RSOS model with the impurity,
\begin{eqnarray}
T(u)=&&W(u\vert \{\sigma_{j+1} \})...
    W(u\vert \{\sigma_{N}\} )
    \widetilde{W}^p(u\vert \{\sigma_{N+1}\} )\nonumber \\
    &&\times W(u\vert \{\sigma_1\})...
      W(u\vert \{\sigma_{j-1}\}),
\label{eqn:trans}
\end{eqnarray}
which could reproduce the
nontrivial properties of the multichannel Kondo  model.
Here $W$'s are the face  weights of the RSOS model with
a spectral parameter $u$,
and $\{\sigma\}$ is the set of spins around the face.
The impurity counterpart of the face weight $\widetilde{W}^p$ is
given by the fusion of $W$ \cite{kp,br,fend,mar,djm},
\begin{equation}
\widetilde{W}^p(u\vert \{\sigma\})
       =\sum_{\{\sigma^{'}\}}\prod_{k=1}^p
         W \biggl( u+\biggl(\frac{p+1}{2}
   -k\biggr)\lambda\biggl\vert\biggr.\{\sigma^{'}\}\biggr),
\end{equation}
where $\lambda=\pi/(n+2)$.
Note that from this equation one can indeed reproduce the
Z$_n$ sector of the thermodynamic Bethe-ansatz
equation obtained for the overscreening case\cite{tw,ad}.

We wish to  study the above  RSOS model with the impurity to
deduce the correct spectrum of the overscreening model, and
show how the fusion hypothesis emerges in this framework.
With a view to  avoiding a pathology in the string hypothesis,
we deal with  (\ref{eqn:trans})
following the functional equation method\cite{res2,kp,kns}.
The essence of this method is that  one can seek
for the solution to the eigenvalue
problem by observing analyticity properties of the
transfer matrix. We shall slightly generalize it to take into
account the effect of the impurity.
Let us first construct the fusion hierarchies of this model
following the standard fusion procedure\cite{kr,djm}.
Introduce first the fused face weight for host
electrons\cite{djm,kp,br},
\begin{equation}
W^{1,q}(u\vert \{\sigma\})
  =\prod_{k=0}^{q-2}s_k^{-1}(u)\sum_{\{\sigma^{'}\}}
 \prod_{k=1}^q W(u+(k-1)\lambda\vert \{\sigma^{'}\} ),
\label{eqn:ffw}
\end{equation}
where
$
s_k(u)=\frac{\sin[u+(k-j)\lambda]}{\sin \lambda},
$
and the fused face weight for the impurity, which
is given by eq.(\ref{eqn:ffw}) with $s_k(u)$ replaced by
$
\tilde{s}_k(u)=\prod_{j=0}^{p-1}
\sin(u+(k+\frac{p-1}{2}-j)\lambda)/
\sin \lambda.
$
We then obtain the transfer matrix for the fusion hierarchies,
\begin{equation}
T^q(u)=W^{1,q}...W^{1,q}\widetilde{W}^{p,q}W^{1,q}...W^{1,q}.
\end{equation}From
the generalized Yang-Baxter equations satisfied by $W^{1,q}$ and
$\widetilde{W}^{p,q}$, it is easily seen that the commutativity of
the transfer matrix holds, $[T^q(u), T^{q^{'}}(v)]=0$,
which guarantees
the construction of the fusion hierarchies.
The fusion hierarchies satisfy the following
functional equations\cite{br,kp},
\begin{eqnarray}
T_0^1T_1^1&=&f_{-1}f_{1}+f_0T_0^2 \\
T_0^qT_q^1&=&f_qT_0^{q-1}+f_{q-1}T_0^{q+1}, \label{eqn:fe}
\end{eqnarray}
where $T_q\equiv T(u+q\lambda)$, and
\begin{equation}
f_q=\biggl(\frac{\sin(u+q\lambda)}{\sin \lambda}\biggr)^N
\prod_{j=0}^{p-1}
\frac{\sin(u+(q+\frac{p-1}{2}-j)\lambda)}{\sin\lambda}.
\end{equation}
  Following Kl\"umper and Pearce\cite{kp}, we recast the
above functional equations into the following form,
\begin{equation}
t_0^qt_1^q=(1+t_1^{q-1})(1+t_0^{q+1}),
\qquad 1\leq q\leq n-1 \label{eqn:fe2}
\end{equation}
with
$t_0^q=T_1^{q-1}T_0^{q+1}/f_{-1}f_q,$
and $t_0^0=t_0^{n}=0$.
If we omit the face weight for the impurity, $\widetilde{W}^p$,
the analyticity strip of $t^q(u)$ is given by\cite{kp}
\begin{equation}
\frac{1-q}{2(n+2)}\pi-\frac{\pi}{2}<
{\rm Re}u< \frac{1-q}{2(n+2)}\pi.
\label{eqn:anad}
\end{equation}
Note that $t^q(u)$ has poles in this strip. For
a large system with linear size $N$,
$t^q(u)$ can be described by
$t^q(u)=(z^q(u))^Ny^q(u)$, where
$(z^q(u))^N$ expresses the host electron part
and $y^q(u)$ the impurity part. From
eq.(\ref{eqn:fe2}), these functions respectively satisfy
\begin{equation}
\frac{z^q(u) z^q(u+\lambda)}
{z^{q-1}(u+\lambda)z^{q+1}(u)}=1,
\label{eqn:z1}
\end{equation}
for $1\leq q \leq n-1$,
with the condition
$z^0(u)=z^n(u)=1$,
and
\begin{equation}
\frac{y_0^qy_1^q}
{y_1^{q-1}y_0^{q+1}}=1, \qquad 1\leq q\leq n-1, \label{eqn:y1}
\end{equation}
with $y^0_0=y^n_0=1$.

Now our task is to find the solution to
$z^q(u)$ and as well as to $y^q(u)$,
which should respect the correct pole structure of $t^q(u)$
in the analyticity strip (\ref{eqn:anad}).
The solution to eq.(\ref{eqn:z1}) was already
found\cite{kp},
\begin{equation}
z^q(u)=\frac{\sin[\frac{n+2}{n}(u)]}
{\sin[\frac{n+2}{n}(u+q\lambda)]},
\end{equation}
which indeed possesses the pole structure correctly.
This immediately yields the well-known  finite-size spectrum of
$Z_n$ parafermion theory\cite{kp},
\begin{equation}
E=\frac{2\pi v}{N}\biggl(\frac{j(j+1)}{n+2}-\frac{m^2}{4n}\biggr),
\label{eqn:zn}
\end{equation}
where $ m=-2j, -2j+2, ..., 2j-2, 2j $, and
$ j=0, \frac{1}{2}, 1, ..., \frac{n}{2} $.
Note that only the holomorphic piece of conformal weights
appears since the Kondo problem is
effectively described by a chiral system\cite{tw,ad}.

In order to obtain the solution to the impurity part, $y^q(u)$,
we have to know how the presence of the
impurity changes analyticity properties of $t^q$.
Although this problem may not be easy in general,
it is found  that for the special case of $p=1$,
the analyticity strip
(\ref{eqn:anad}) is slightly modified due to the impurity. Hence
the unique solution to $y^q$ can be easily obtained,
which reproduces the pole structure of $t^q$ under the
condition $y^0_0=y^{n}_0=1$. We find
\begin{equation}
y^q(u)=\prod^{p-1}_{l=0}
\frac{\sin[\frac{2(n+2)}{n}(u+(\frac{p-1}{2}-l)\lambda)]}
{\sin[\frac{2(n+2)}{n}(u+(q+\frac{p-1}{2}-l)\lambda)]}.
\label{eqn:yq}
\end{equation}
In the following we will restrict our
discussions to the case $p=1$, owing to the above
technical reason.

Let us now turn to the calculation of the finite-size spectrum.
To this end, it is convenient to introduce the new functions,
$a^q(x)\equiv  [t^q(\frac{n}{n+2}ix
-(q+\frac{n}{2})\frac{\lambda}{2})]^{-1}$.
Applying the method of Kl\"umper and Pearce\cite{kp},
we obtain the non-linear integral equation for $a^q$,
\begin{equation}
\ln a^q=\ln e^q+\sum_r K_{q,r}*\ln (1+a^r)+D^q, \label{eqn:lae}
\end{equation}
where the asterisk stands for the convolution.
Here $D^q$ is the vector independent of $x$
(see eq.(\ref{eqn:dq}))
and Fourier components of the matrix $K$ are given by
the solution of the equation,
\begin{equation}
(K_0(z)\delta_{m,l}-\delta_{m, l-1}-\delta_{m, l+1})K_{l, n}
=-\delta_{m, n+1}-\delta_{m, n-1},
\end{equation}
with $K_0(z)=-2\cosh(\pi z/n)$.
$\ln e^q$ is determined by $t^q=(z^q(u))^Ny^q(u)$, of which
the bulk part is given in ref.\cite{kp}.
The impurity part is given by
\begin{equation}
\sum_{j=0}^{p-1}\ln
\biggl(\frac{\sinh\{x-[\frac{q}{2}-
\frac{n}{4}+\frac{p-1}{2}-j]\frac{2\pi i}{n}\}}
{\sinh\{x
-[-\frac{q}{2}-\frac{n}{4}+\frac{p-1}{2}-j]
\frac{2\pi i}{n}\}}\biggr),
\end{equation}from which one can easily obtain the scaling limit
of $\ln e^q$,
\begin{equation}
\lim_{N\rightarrow\infty}\ln e^q(\pm (x+\ln N))=
(\mbox{bulk part})\mp\frac{2i\pi pq}{n}.\label{eqn:les}
\end{equation}

We are now ready to compute the finite-size
spectrum from the finite-size corrections
to the eigenvalue of the transfer matrix,
\begin{equation}
b^q(x)\equiv T^q_{\rm finite}\biggl(\frac{n}{n+2}ix
-\biggl(q-1+\frac{n}{2}\biggr)\frac{\lambda}{2}\biggr),
\end{equation}
where $T_{\rm finite}^q$ is the finite-size corrections
to the transfer matrix.
The integral equation for $b^q(x)$ is easily found\cite{kp},
\begin{equation}
\ln b^q=\sum_r \hat{K}_{q,r}*\ln (1+a^r)+C^q, \label{eqn:lbe}
\end{equation}
where the kernel $\hat{K}$ is given by
\begin{equation}
\hat{K}_{j,l}=\frac{\sin\frac{j\pi}{n}\sin\frac{l\pi}{n}}
{\pi\sin\frac{\pi}{n}}e^{-\vert x\vert},
\end{equation}
and $C^q$ is the vector consisting of constant elements
which do not contain $1/N$ corrections. From eqs.(\ref{eqn:lae})
and (\ref{eqn:lbe}), one can obtain
the finite-size spectrum\cite{kp},
\begin{eqnarray}
\frac{N\pi E_{RSOS}}{v}&=&\frac{N\pi (\ln b^1 -C^1)}{v}
\nonumber \\
&=&\sum_{q=1}^{n-1}L\biggl(\frac{a^q_{\infty}}
{A^q_{\infty}}\biggr)
+\frac{1}{2}\sum_{q=1}^{n-1}D^q \ln A^q_{\infty},
\label{eqn:enes}
\end{eqnarray}
where $L(x)$ is the Rogers dilogarithm.
Here $a^q_{\infty}=\lim_{x\rightarrow \infty}a^q(x)$, and
$A^q_{\infty}=\lim_{x\rightarrow \infty}(1+a^q(x))$,
which are evaluated from the asymptotics of
eq.(\ref{eqn:fe2}) as\cite{kp},
\begin{equation}
a^q_{\infty}=\frac{\sin^2\theta}{\sin q\theta\sin(q+2)\theta},
\quad
A^q_{\infty}=
\frac{\sin^2(q+1)\theta}{\sin q\theta\sin(q+2)\theta},
\end{equation}
where $\theta$ is a multiple of $\pi/(n+2)$.
$D^q$ is determined from the asymptotic
behavior of eq.(\ref{eqn:lae}),
\begin{equation}
D^q=(\mbox{bulk part})
- \frac{2pq\pi i}{n}. \label{eqn:dq}
\end{equation}
The second term comes from the
 impurity contribution,
and is responsible for  determining the selection
rule for quantum numbers, as we will see shortly.

We thus find from eqs.(\ref{eqn:enes}) and (\ref{eqn:dq})
the correction of
the finite-size spectrum due to the impurity,
\begin{eqnarray}
\Delta E_{RSOS}&=& -\frac{i p}{nN}\ln\biggl(\frac{\sin^{n} n\theta}
{\sin\theta \sin^{n-1}(n+1)\theta}\biggr)  \nonumber \\
&=& \frac{i p}{nN}\biggl(i\pi m
+n\ln\biggl(2\cos\frac{\pi s}{n+2}\biggr)\biggr), \label{eqn:desos}
\end{eqnarray}
Here $s=1, 2, ..., n+1$, and $m=s-1$ (mod 2),
$\vert m \vert \leq s-1=2j$.
The second term in the bracket can be evaluated from
the braid limit\cite{kp,pea},
\begin{equation}
\lim_{{\rm Im}u\rightarrow \pm \infty}
\biggl(\frac{\sin\lambda}{\sin(u-\lambda/2)}\biggr)^N
T(u)=2\cos\frac{\pi s}{n+2}. \label{eqn:brl}
\end{equation}
Since the left-hand side of eq.(\ref{eqn:brl}) is evaluated as
$i\pi m/n$,
the second term in the bracket of eq.(\ref{eqn:desos}) is
equal to $i\pi m$.
Consequently  we end up with the simple formula,
\begin{equation}
\Delta E_{RSOS}=-\frac{2\pi}{nN}mp.\label{eqn:de}
\end{equation}
This completes the calculation of the finite-size corrections
due to the impurity.

Now combining  the ordinary
finite-size spectrum for $Z_n$ parafermions
eq.(\ref{eqn:zn}) and the impurity contribution
eq.(\ref{eqn:de}), we obtain the desired result,
\begin{equation}
\frac{NE_{RSOS}}{2\pi v}= \frac{j(j+1)}{n+2}-\frac{(m+p)^2}{4n}
+\mbox{const},
\label{eqn:fsrsos}
\end{equation}
where $ m=2j\quad(\mbox{mod} 2)$, and
$j=0, 1/2, 1, ..., n/2$.
One can see that the spectrum fits in with
$Z_n$ parafermion theory, and {\it only the selection rule for
quantum numbers is changed via $m \rightarrow (m+p)$,
after incorporating the impurity effect}.
We think that this may be a microscopic description of the fusion
hypothesis.

To confront our results with those of Affleck and
Ludwig\cite{la}, let us look at the total finite-size spectrum,
which is given by the sum of
eqs.(\ref{eqn:fss1}) and (\ref{eqn:fsrsos}).
One can easily see that the term  $-(m+p)^2/4n$
in (\ref{eqn:fsrsos})
should be canceled by  $S_z^2/n$ in (\ref{eqn:fss1})
by suitably choosing the quantum number for $S_z$.
This in turn modifies the quantum numbers in
SU(2)$_n$ Kac-Moody algebra. As a result, the total spectrum
is reduced to
\begin{eqnarray}
\frac{NE}{2\pi v}&=&\frac{(Q-n)^2}{4n}
+\frac{\tilde j(\tilde j+1)}{n+2}
+(\mbox{flavor part}) \nonumber \\
&&+n_Q+n_s+n_f, \label{eqn:fss2}
\end{eqnarray}
with newly introduced quantum numbers
$\tilde j=\vert j-p/2\vert$ where $\tilde
j=0, 1/2, 1, ... , n/2$. Note that
$Q$ and $j$ are the charge and spin  quantum
numbers for free electrons without Kondo impurities.
Consequently we can say that the sole effect
of the Kondo impurity is to modify the selection rule for
quantum numbers in the  spin spectrum, via
$j \rightarrow \tilde j$,
which indeed leads to non-Fermi liquid properties for
the overscreening Kondo effect.
This is the essence of
the Kac-Moody fusion hypothesis proposed
by Affleck and Ludwig\cite{la}.

In this paper we have been concerned with the special case
$p=(n-2S)=1$ in the overscreening model.
As mentioned above, we have encountered some
technical problems  for treating the case with
general $p$, which have not been resolved yet.
Nevertheless we think that the solution
to $y^q$ for general $p$ is given by eq.(\ref{eqn:yq}), and
the corresponding finite-size spectrum
should take the form of eq.(\ref{eqn:fss2}).
We hope to clarify these points in the future work
to establish the present conclusion for the general case.

The authors are grateful to A. W. W. Ludwig
and A. Kl\"umper for illuminating discussions.
This work was partly supported by a Grant-in-Aid from the Ministry
of Education, Science and Culture.



\end{multicols}

\begin{references}
\bibitem{multi} P. Nozieres and A. Blandin, J. Phys. (Paris)
{\bf 41}, 193 (1980).

\bibitem{tw} A. M. Tsvelick and P. B. Wiegmann,
Z. Phys. {\bf B54}, 201 (1985); J. Stat. Phys. {\bf 38},
125 (1985).

\bibitem{ad} Andrei and Destri, Phys. Rev. Lett. {\bf 52},
364 (1984).

\bibitem{sch} P. Schlottmann and P. D. Sacramento,
Adv. Phys. {\bf 42}, 641 (1993).

\bibitem{multi2} D. L. Cox, Phys. Rev. Lett. {\bf 59}, 1240 (1987).

\bibitem{ks} M. Koga and H. Shiba, preprint.

\bibitem{la} I. Affleck and A. W. W. Ludwig, Nucl. Phys.
{\bf B360}, 641
(1991); {\bf B352}, 849 (1991);
Phys. Rev. Lett. {\bf 67}, 161 (1991);
Phys. Rev. {\bf B48}, 7279 (1993);
A. W. W. Ludwig and I. Affleck, Phys. Rev. Lett. {\bf 67}
(1991) 3160; Nucl. Phys. {\bf B428}, 545 (1994).

\bibitem{exact}
N. Andrei, K. Furuya and J. H. Lowenstein,
Rev. Mod. Phys. {\bf 55}, 331 (1983).

\bibitem{tw2}
A. M. Tsvelick and  P. B. Wiegmann, Adv. Phys.
{\bf 32}, 453 (1983).

\bibitem{fks} S. Fujimoto, N. Kawakami, and S. -K. Yang, Phys. Rev.
{\bf B50}, 1046 (1994).

\bibitem{res2} N. Yu. Reshetikhin, Lett. Math. Phys. {\bf 7},
205 (1983).

\bibitem{kp} A. Kl\"umper and P. A. Pearce,
Physica {\bf A183}, 304 (1992).

\bibitem{kns} A. Kuniba, T. Nakanishi, and J. Suzuki,
Int. J. Mod. Phys. {\bf A9}, 5215, 5267 (1994).

\bibitem{dw} H. J. de Vega and F. Woynarovich, Nucl. Phys.
{\bf B251}, 439 (1985).

\bibitem{woy} F. Woynarovich, J. Phys. {\bf A22}, 4243 (1989).

\bibitem{am}
F. C. Alcaraz and M. J. Martins, J. Phys. {\bf A22},
1829 (1989).

\bibitem{res} N. Yu. Reshetikhin, J. Phys. {\bf A24}, 3299 (1991).

\bibitem{zz} A. B. Zamolodchikov and Al. B. Zamolodchikov,
Nucl. Phys. {\bf B379}, 602 (1992).

\bibitem{fend} P. Fendley, Phys. Rev. Lett. {\bf 71}, 2485 (1993).

\bibitem{abf} G. E. Andrews, R. J. Baxter, and P. J. Forrester,
J. Stat. Phys. {\bf 35}, 193 (1984).

\bibitem{zf} A. B. Zamolodchikov and V. V. Fateev, Sov. Phys. JETP
{\bf 62}, 215 (1985).

\bibitem{ts} A. M. Tsvelick, Nucl. Phys. {\bf B305}, 675 (1988).

\bibitem{mar} M. J. Martins, Nucl. Phys. {\bf B426}, 661 (1994).


\bibitem{br} V. V. Bazhanov and N. Reshetikhin,
Int. J. Mod. Phys. {\bf A4},
115 (1989).

\bibitem{djm} E. Date, M. Jimbo, T. Miwa, and M. Okado,
Lett. Math. Phys. {\bf 12}, 209 (1986).

\bibitem{kr} P. P. Kulish, N. Yu. Reshetikhin, E. K. Sklyanin,
Lett. Math. Phys. {\bf 5}, 393 (1981).

\bibitem{pea} P. A. Pearce, in Proc. RIMS91, Infinite Analysis
(World Scientific).

\end{references}
\end{document}